\documentclass[11pt]{article}
\usepackage[T1]{fontenc}
\usepackage[utf8]{inputenc}
\usepackage{amsmath,newtxtext}
\usepackage[notext]{stix2}
\usepackage[a4paper,margin=22mm]{geometry}
\usepackage{graphicx,booktabs,array}
\usepackage{microtype}
\usepackage[authoryear,round]{natbib}
\usepackage{xurl}
\usepackage[hidelinks,unicode]{hyperref}
\providecommand{\doi}[1]{doi: \href{https://doi.org/#1}{\nolinkurl{#1}}}
\DeclareUnicodeCharacter{00B2}{\textsuperscript{2}}
\DeclareUnicodeCharacter{2074}{\textsuperscript{4}}
\DeclareUnicodeCharacter{2076}{\textsuperscript{6}}
\DeclareUnicodeCharacter{0394}{\ensuremath{\Delta}}
\DeclareUnicodeCharacter{03A6}{\ensuremath{\Phi}}
\DeclareUnicodeCharacter{03B1}{\ensuremath{\alpha}}
\DeclareUnicodeCharacter{03B3}{\ensuremath{\gamma}}
\DeclareUnicodeCharacter{03B4}{\ensuremath{\delta}}
\DeclareUnicodeCharacter{03C1}{\ensuremath{\rho}}
\DeclareUnicodeCharacter{00D7}{\ensuremath{\times}}
\DeclareUnicodeCharacter{00B1}{\ensuremath{\pm}}
\DeclareUnicodeCharacter{2010}{-}
\DeclareUnicodeCharacter{2011}{-}
\DeclareUnicodeCharacter{2192}{\ensuremath{\rightarrow}}
\DeclareUnicodeCharacter{2194}{\ensuremath{\leftrightarrow}}
\DeclareUnicodeCharacter{2202}{\ensuremath{\partial}}
\DeclareUnicodeCharacter{2212}{\ensuremath{-}}
\DeclareUnicodeCharacter{223C}{\ensuremath{\sim}}
\DeclareUnicodeCharacter{2248}{\ensuremath{\approx}}
\DeclareUnicodeCharacter{2261}{\ensuremath{\equiv}}
\DeclareUnicodeCharacter{2264}{\ensuremath{\leq}}
\DeclareUnicodeCharacter{226A}{\ensuremath{\ll}}
\DeclareUnicodeCharacter{226B}{\ensuremath{\gg}}
\DeclareUnicodeCharacter{27E8}{\ensuremath{\langle}}
\DeclareUnicodeCharacter{27E9}{\ensuremath{\rangle}}

\begin{document}
\begingroup
\setlength{\parindent}{0pt}
\setlength{\parskip}{6pt}
\begin{center}

{\LARGE \textbf{Elasticity of polycrystalline davemaoite constrains its grain size in the lower mantle}\par}
\vspace{5pt}


Peiyu Zhang\textsuperscript{1}, Liang Yuan\textsuperscript{1,2}, Rong Huang\textsuperscript{3}, Yu Ye\textsuperscript{1}, Xiang Wu\textsuperscript{1}, Junfeng Zhang\textsuperscript{1}

{\small

\textsuperscript{1}State Key Laboratory of Geological Processes and Mineral Resources, School of Earth and Planetary Sciences, China University of Geosciences, Wuhan, China;


\textsuperscript{2}Bayerisches Geoinstitut, Universität Bayreuth, Bayreuth, Germany;


\textsuperscript{3}Earth and Environmental Sciences, University of Michigan, Ann Arbor, USA.

}

Correspondence: liang.yuan@uni-bayreuth.de

\end{center}

\textbf{Key points}


\begin{itemize}
\setlength{\itemsep}{2pt}
\setlength{\parsep}{0pt}
\setlength{\parskip}{0pt}
\item
  Million-atom machine-learning simulations show that 5.8 vol\% disorder lowers davemaoite shear and bulk moduli by 37\% and 12\%.
\item
  Nanoscale grain-boundary disorder reconciles the longstanding \textasciitilde30\% shear modulus gap between theory and experiments.
\item
  Seismic constraints on disorder--elasticity scaling yield a minimum lower-mantle davemaoite grain size of \textasciitilde100 nm.
\end{itemize}




\textbf{Abstract}\par
Earth's lower mantle hosts large low shear-velocity provinces (LLSVPs) beneath Africa and the Pacific, but their origin remains debated. Davemaoite, a major lower-mantle mineral, has been proposed as a contributor to these anomalies; however, its experimentally measured shear modulus is \textasciitilde30\% lower than first-principles predictions---a longstanding discrepancy between theory and experiment. Here, using million-atom machine-learning molecular dynamics simulations, we show that nanoscale grain-boundary disorder, invisible to conventional x-ray diffraction, can reconcile this gap. Introducing 5.8 vol\% disordered regions reduces the shear and bulk moduli by 37\% and 12\%, respectively. This shear-selective softening can reproduce LLSVP-like seismic anomalies in basalt-rich assemblages, but only at disorder levels corresponding to grain sizes smaller than those expected in the lower mantle. Combining the disorder--elasticity relationship with seismic constraints, we derive a minimum davemaoite grain size of \textasciitilde100 nm. This limit is consistent with mantle grain growth models and with attenuation-based evidence for coarse-grained LLSVPs.


\textbf{Plain language summary}


How large are mineral grains in Earth's deep mantle? This basic property is difficult to measure directly, because samples from these depths are extremely rare and their original textures are usually altered on their way to the surface. Here we show that grain size can instead be constrained indirectly, from the elastic behavior of davemaoite, a major lower-mantle mineral. Computer simulations tracking millions of atoms reveal that disordered grain boundaries strongly reduce the mineral's resistance to shearing. By comparing this softening effect with seismic observations of Earth's largest low-velocity regions, we estimate that davemaoite grains must be at least about 100 nanometers in size. This limit agrees with independent evidence from seismic attenuation, supporting the view that these vast regions are composed of relatively coarse-grained, long-lived mantle materials.

\par
\endgroup
\clearpage
\section*{1. Introduction}


Two large low shear-velocity provinces (LLSVPs) beneath the Pacific Ocean and Africa constitute the most prominent seismic heterogeneities in Earth's lower mantle. Rising more than 1,000 km from the core--mantle boundary and covering nearly one-quarter of its area, these structures are of debated origin, with thermal, compositional, and combined thermochemical models all proposed (\citealp{Garnero2016}). Global 3D attenuation models built from whole-Earth oscillations implicate grain size as a potential contributor to LLSVP seismic signatures: coarse-grained LLSVPs appear surrounded by finer-grained circum-Pacific mantle (\citealp{TalaveraSoza2025}). Davemaoite (CaSiO\textsubscript{3} perovskite; Dvm)\footnote{We follow IMA nomenclature (IMA2020-012a; \citealp{Tschauner2021}) for CaSiO\textsubscript{3} perovskite (davemaoite), although its reported occurrence as a diamond inclusion remains debated (\citealp{Walter2022}).} is a leading candidate for explaining these lower-mantle velocity anomalies: it is the third most abundant lower-mantle phase, constituting \textasciitilde10 vol\% of pyrolite and up to \textasciitilde25 vol\% of subducted basalt (\citealp{Greaux2019}; \citealp{Thomson2019}; \citealp{Zhou2025}).


Assessing Dvm's contribution to LLSVPs---including both their velocity and attenuation signatures---is hindered by two fundamental uncertainties. First, grain size, a first-order control on attenuation (\citealp{Karato2008}), remains largely unknown because direct samples of the lower mantle are very sparse. Second, the elasticity of Dvm under lower-mantle conditions remains controversial. Dvm cannot be quenched to ambient conditions: it undergoes a cubic-to-tetragonal phase transition during cooling (\citealp{Ono2004}) and amorphizes upon decompression (\citealp{Irifune1994}), so its elastic properties must be measured \emph{in situ} at high pressures (\emph{P}) and temperatures (\emph{T}). Even so, reported sound velocities differ by \textasciitilde10\% between experimental studies (\citealp{Greaux2019}; \citealp{Thomson2019}). More critically, a longstanding gap separates experiment and theory: density functional theory (DFT) calculations consistently overestimate the shear modulus of Dvm by \textasciitilde30\% relative to experimental measurements (\citealp{Kawai2015}; \citealp{Stixrude2007}). This discrepancy is exceptional among major mantle minerals, for which theory and experiment generally agree closely (\citealp{Tsuchiya2020}), and has remained unresolved for decades.


A likely explanation lies in microstructure: elasticity is sensitive to grain boundary (GB) disorder, which increases as grain size decreases (\citealp{Marquardt2020}). GBs in polycrystals are mechanically compliant, localize strain (\citealp{Karato2008}), and can promote amorphization (\citealp{Irifune1994}). Although Dvm microstructures cannot be preserved during recovery, \emph{ex situ} studies of mantle silicate aggregates show that intergranular amorphous films are common (\citealp{Samae2021}). Such compliant interfacial phases reduce aggregate elastic moduli in materials as diverse as stishovite (\citealp{Buchen2018}), periclase (\citealp{Gleason2011}), and ice (\citealp{Sayers2018}).


To date, all high \emph{P--T} velocity measurements of Dvm have necessarily been performed on polycrystalline samples (p-Dvm), typically synthesized \emph{in situ} from glass precursors using large volume presses (LVPs; \citealp{Greaux2019}) or diamond anvil cells ( DACs; \citealp{Zhou2025}), rather than on single crystals, which would isolate the intrinsic elasticity of Dvm from microstructural effects. Critical evaluation of glass-derived synthesis routes indicates that, in the davemaoite--bridgmanite system, glassy starting materials impose kinetic limitations that yield nanocrystalline products with grain sizes of only tens of nanometers and abundant GBs (\citealp{Zhang2026}). \emph{In situ} synchrotron x-ray diffraction (XRD) is routinely used to assess sample crystallinity before elasticity measurements, but it is insensitive to small fractions of amorphous or GB-disordered material that may persist even after high-\emph{T} annealing (\citealp{Choi2004}): such nanoscale disorder generates only weak diffuse scattering that can be readily obscured by Bragg reflections from the crystalline matrix (\citealp{Egami2003}; \citealp{Klug1974}; \citealp{Smith2018}). Consequently, existing velocity measurements inevitably include the elastic effects of GB disorder and potentially untransformed amorphous starting material, yet the abundance of this disorder and its contributions to compressional- and shear-wave velocities (\emph{V}\textsubscript{P} and \emph{V}\textsubscript{S}) remain unconstrained. Whether nanoscale GB disorder can explain the experiment--theory discrepancy is unknown---and whether this sensitivity could, in principle, be inverted to constrain lower-mantle grain size has not been examined.


In this work, we use large-scale molecular dynamics (MD) simulations driven by DFT-based machine-learning interatomic potentials (MLIPs) to investigate the role of GB disorder in Dvm elasticity. Million-atom simulations of Dvm polycrystals explicitly resolve GB networks under lower-mantle conditions and quantify their influence on aggregate elastic properties and seismic velocities. We show that nanoscale GB disorder substantially reduces the shear modulus (and thus \emph{V}\textsubscript{S}) of p-Dvm, and we establish a quantitative relationship between elastic softening and disorder fraction. Comparing this relationship with seismic constraints yields a lower bound on the grain size of lower-mantle Dvm. These results provide a mineral physics constraint on lower-mantle grain size that complements independent estimates from seismic attenuation.


\section*{2. Methods}


DFT is the standard tool for mineral elasticity at lower-mantle conditions (\citealp{Tsuchiya2020}), but its cubic computational scaling confines calculations to small systems---a limitation that is particularly severe for Dvm, whose elastic properties converge unusually slowly with system size (\citealp{Kawai2015}). MLIPs (e.g., \citealp{Behler2007}) overcome this limitation, reproducing DFT-level accuracy at far lower cost and enabling the large-scale MD needed to resolve GB networks in p-Dvm.


\emph{MLIP training.} We built MLIPs for CaSiO\textsubscript{3} using the Deep Potential method (\citealp{Wang2018}), following the training procedure in our previous work (\citealp{Yuan2023}; \citealp{ZhangP2025}). A central challenge in MLIP development is balancing dataset compactness with the diversity needed for accuracy and transferability. We addressed this using two complementary active learning strategies: (i) DP-GEN (\citealp{Zhang2020}) to sample simple solid/liquid configurations (\textasciitilde10\textsuperscript{2} atoms), and (ii) an in-house workflow targeting the complex GB local environments in p-Dvm (\textasciitilde10\textsuperscript{5} atoms). Together, these strategies sampled configurations spanning 0--100 GPa and 0--6,000 K. We then optimized the final MLIP on this dataset using the Neuroevolution Potential (NEP) framework (\citealp{Fan2021}), which leverages GPU parallelism for a throughput of \textasciitilde10\textsuperscript{7} atom-steps s\textsuperscript{-1}---an order of magnitude faster than typical MLIP implementations. The NEP model reproduces both DFT calculations and experimental measurements, resolving discrepancies in previous DFT results (e.g., the \emph{c}/\emph{a} axial ratios).


\emph{MD simulations.} MLIP-driven MD simulations for elasticity calculations were performed using GPUMD (\citealp{Fan2022}). Systems were first equilibrated at target \emph{P--T} conditions in the isothermal--isobaric (\emph{NPT}) ensemble, then evolved in the isothermal--isochoric (\emph{NVT}) ensemble to compute thermoelastic properties. Pressure was controlled with a stochastic cell rescaling barostat (\citealp{Bernetti2020}) and temperature with the Bussi--Donadio--Parrinello thermostat (\citealp{Bussi2007}), with a timestep of 2.0 fs.


\emph{DFT settings.} Single-point DFT calculations for MLIP development were performed in VASP (\citealp{Kresse1996}) using the projector augmented-wave method (\citealp{Kresse1999}) and the SCAN meta-generalized gradient approximation (\citealp{Sun2015}). SCAN has been shown to outperform LDA (\citealp{Kohn1965}), PBE (\citealp{Perdew1996}), and PBEsol (\citealp{Perdew2008}) for thermoelastic properties of MgSiO\textsubscript{3}--CaSiO\textsubscript{3}--SiO\textsubscript{2} systems (\citealp{Wang2026}). A plane-wave cutoff energy of 800 eV was used, corresponding to twice the maximum ENMAX among the pseudopotentials. The valence electron configurations were Ca (3\emph{s}²3\emph{p}⁶4\emph{s}²), Si (3\emph{s}²3\emph{p}²), and O (2\emph{s}²2\emph{p}⁴). Self-consistent field calculations were converged to 10\textsuperscript{-6} eV. Brillouin zone sampling was controlled via KSPACING = 0.2 Å\textsuperscript{-1}, ensuring consistent \emph{k}-point density across the varied cell geometries in the training set; for a representative 120-atom CaSiO\textsubscript{3} cell, this yielded a 3 × 3 × 3 \emph{k}-point grid.


\section*{3. Results}


\section*{3.1. Phase stability}


The stable structure of CaSiO\textsubscript{3} perovskite under lower-mantle conditions remains debated. Although cubic symmetry is widely assumed, both experiments and theory predict low-\emph{T} tetragonal distortions (\citealp{Komabayashi2007}; \citealp{Ono2004}; \citealp{Stixrude1996}). Deviatoric stresses, which stabilize the tetragonal phase, further complicate experimental constraints on the tetragonal--cubic transition (\citealp{Chen2018}). Theoretical predictions of the transition \emph{T} span 500--1,500 K at 50 GPa (\citealp{Thomson2019}; \citealp{Wu2024}; \citealp{ZhangC2025}). Identifying the stable phase at relevant \emph{P--T} conditions is therefore a prerequisite for self-consistent elasticity calculations.


We assessed phase stability by computing free energy differences using nonequilibrium thermodynamic integration (NE-TI; \citealp{Freitas2016}). Unlike approaches that fix \emph{c}/\emph{a} at a reference value (e.g., \emph{c}/\emph{a} = 1.01; \citealp{Thomson2019}), we allowed the lattice ratio \emph{c}/\emph{a} to evolve freely, consistent with experimental evidence that it varies systematically with \emph{P} and \emph{T} (\citealp{Sun2022}). Lattice constants were determined from a 5,000-atom supercell (10 × 10 × 10 unit cells) equilibrated in the \emph{NPT} ensemble for 440 ps; free energy differences were then obtained from forward and reverse NE-TI simulations of 1 ns each.


Figure 1 shows the Gibbs free energy, \emph{G}(\emph{c}/\emph{a, P, T}), across \emph{c}/\emph{a} = 0.995--1.006. Unlike 0-K DFT calculations, which capture only enthalpic contributions and predict sharp energy differences between competing phases (\citealp{Caracas2005}), our NE-TI calculations account for anharmonic lattice vibrations, yielding a nearly flat free energy landscape. Run-to-run fluctuations in individual NE-TI calculations at fixed \emph{c}/\emph{a} (\textasciitilde120 × 10\textsuperscript{-6} eV atom\textsuperscript{-1}) exceed the variation in the ensemble-averaged free energy (\(\overline{G}\)) across the entire \emph{c}/\emph{a} range (\textasciitilde20 × 10\textsuperscript{-6} eV atom\textsuperscript{-1}). Both quantities are orders of magnitude smaller than the typical \textasciitilde10\textsuperscript{-3} eV atom\textsuperscript{-1} precision of DFT-MD TI (\citealp{Alfe2003}). Resolving such subtle \(\overline{G}\) differences requires hundreds of independent calculations per \emph{c}/\emph{a} value, up to 300 at the most sensitive conditions. We therefore restricted calculations to 50 and 12 GPa, where well-constrained experimental \emph{c}/\emph{a} data for Dvm minimize uncertainties arising from deviatoric stress (\citealp{Chen2018}). Our calculations do not extend to the megabar \emph{P} at the core--mantle boundary, which approach or exceed the MLIP's training range (0--100 GPa) and thus risk degraded accuracy. Extending this analysis to the full lower-mantle \emph{P} range (via additional NE-TI sampling and possibly MLIP retraining over a wider \emph{P--T} range) is straightforward in principle but computationally prohibitive.


At 300~K (50 GPa), the \(\overline{G}\) profile exhibits two minima (Fig.~1d). The global \(\overline{G}\) minimum at \emph{c}/\emph{a}~∼~1.004 corresponds to a slightly elongated tetragonal phase; a metastable contracted state at \emph{c}/\emph{a}~∼~0.998 lies marginally higher in \(\overline{G}\). The predicted distortion (\emph{c}/\emph{a}~∼~1.004) is more subtle than earlier DFT results (\emph{c}/\emph{a} = 1.02; \citealp{Stixrude2007}) but agrees closely with stress-minimized DAC measurements (\emph{c}/\emph{a}~∼~1.005; \citealp{Chen2018}). Experimental studies also report metastable contracted states with \emph{c}/\emph{a} = 0.995--0.998 (\citealp{Ono2004}; \citealp{Shim2002}). With increasing \emph{T}, the equilibrium \emph{c}/\emph{a} approaches unity: a slight elongation persists at 350~K, but the \(\overline{G}\) minimum reaches \emph{c}/\emph{a} = 1 by 400~K. Pressure reinforces the distortion, shifting the tetragonal--cubic phase transition from 250--300 K at 12 GPa to 350--400 K at 50 GPa. The resulting phase boundary, derived here from unconstrained \emph{c}/\emph{a} free energy calculations, falls at the low-\emph{T} end of the ranges reported by previous studies based on fixed \emph{c}/\emph{a} assumptions.


At lower-mantle conditions (50 GPa, \textasciitilde2,142 K along the 1,600-K adiabat of \citealp{Brown1981}), the \(\overline{G}\) profile remains flat around \emph{c}/\emph{a} = 1, with elongated (\emph{c}/\emph{a} = 1.005) and contracted (\emph{c}/\emph{a} = 0.995) tetragonal distortions differing from the cubic structure by only \(\Delta\overline{G}_{\mathrm{p.f.u.}}/k_{\mathrm{B}}T \approx 4\times10^{-3}\). These nearly degenerate states have indistinguishable Boltzmann weights, driving rapid fluctuations among energetically equivalent tetragonal variants. Because the distortions are symmetrically distributed about \emph{c}/\emph{a} = 1, the ensemble average is ⟨\emph{c}/\emph{a}⟩ = 1, so macroscopic XRD records an apparently cubic structure despite instantaneous local tetragonal distortions.


\section*{3.2. p-Dvm: structure and dynamics}


To model p-Dvm, we constructed simulation cells of \textasciitilde300 × 300 × 300 Å\textsuperscript{3} containing 15, 120, and 480 randomly oriented grains via Voronoi tessellation implemented in Atomsk (\citealp{Hirel2015}). Grains were seeded from a 5-atom cubic CaSiO\textsubscript{3} unit cell equilibrated at 50 GPa and 2,142 K. Geometric truncation at GBs disrupts stoichiometric CaSiO\textsubscript{3} units, producing local Ca:Si:O imbalances; we corrected these iteratively by removing atoms from locally overrepresented species, with \emph{NPT} relaxation and neighbor-list updates after each step, until stoichiometry converged locally and globally. The final p-Dvm structures (3.3--3.4 × 10\textsuperscript{6} atoms) were annealed through repeated heating--quenching cycles between 5,500 and 2,142 K.


To quantify GB volume fraction \emph{Φ}, each structure was partitioned into crystalline grain interiors and disordered regions comprising GB films and isolated amorphous pockets. We identified crystalline and disordered atoms using two complementary descriptors: a structural measure (\emph{s}\textsubscript{2}) and a dynamical measure (atomic displacement). We quantified structural disorder using the two-body entropy, \emph{s}\textsubscript{2} (Fig. 2a; \citealp{Piaggi2017}). Crystalline environments exhibit sharp, well-defined coordination shells and strongly negative \emph{s}\textsubscript{2} values, whereas structurally disordered regions produce broadened local correlations and elevated \emph{s}\textsubscript{2}. The \emph{s}\textsubscript{2} distributions show two populations, with crystalline and disordered populations separated near -5.6 \emph{k}\textsubscript{B}. The low-entropy peak (-7 to -8 \emph{k}\textsubscript{B}) corresponds to ordered grain interiors, whereas the higher-entropy peak (-5 to -4 \emph{k}\textsubscript{B}) marks disordered GBs and isolated amorphous pockets. Two-dimensional \emph{s}\textsubscript{2} maps reveal an evolution in GB topology: in the 15-grain model, disorder is confined to thin intergranular films surrounding large crystalline domains, whereas the 120- and 480-grain aggregates develop progressively denser and more interconnected GB networks.


To characterize GB dynamics, we computed atomic displacement distributions (ADDs) after 1 ns at 2,142 K. Owing to the orders-of-magnitude contrast between GB and intracrystalline self-diffusivities, the ADDs are distinctly bimodal (Fig. 2b). A dominant peak near 0.25 Å reflects thermal vibrations within crystalline grain interiors and closely matches the ADD of a Dvm single crystal at identical \emph{P--T} conditions. In contrast, a broad tail emerges beyond \textasciitilde1.1 Å, indicating diffusive rather than vibrational motion. Species-dependent broadening follows established diffusivity trends in silicate melts (\(D_{\mathrm{Si}}^{\mathrm{GB}} \approx D_{\mathrm{Ca}}^{\mathrm{GB}} < D_{\mathrm{O}}^{\mathrm{GB}}\); \citealp{DeKoker2010}). The bimodal ADD therefore provides a dynamical criterion for separating crystalline and disordered populations. We define the reference threshold \emph{d}\textsubscript{0}---the maximum vibrational amplitude about a lattice site---as the upper bound of the single-crystal ADD. Atoms in p-Dvm with displacements \textgreater{}\emph{d}\textsubscript{0} are classified as disordered; all others are assigned to crystalline grain interiors.


The structural and dynamical classifications capture complementary aspects of GB disorder: \emph{s}\textsubscript{2} identifies locally distorted environments, whereas the atom resolved ADD detects atoms undergoing non-vibrational, diffusive motion. In amorphous materials, disordered bonding prevents local forces from fully canceling under strain, generating residual forces that are relaxed through nonaffine atomic displacements (\citealp{Zaccone2014}). This effect is strongly shear-sensitive: nonaffine relaxations disrupt force cancellation under shear, substantially reducing the shear modulus, whereas compression remains dominated by bond-aligned deformation and largely preserves the bulk modulus. Because \emph{s}\textsubscript{2} cannot identify which distorted atoms undergo such relaxations, we adopt ADD following \citet{Mantisi2017} to quantify disorder fractions of \emph{Φ} = 0\%, 0.9 at\% (1.8 vol\%), 1.8 at\% (3.5 vol\%), and 3.2 at\% (5.8 vol\%) for the single-crystal, 15-, 120-, and 480-grain models, respectively.


\section*{3.3. Single-crystal elasticity}


Because the computed tetragonal--cubic phase boundary for Dvm falls below the coldest slab geotherms, the cubic phase is stable throughout the lower mantle. We therefore compute single-crystal elastic constants (\emph{C\textsubscript{ij}}) and seismic velocities (\emph{V}\textsubscript{P}, \emph{V}\textsubscript{S}) for this phase to establish a reference baseline for the polycrystalline analysis that follows (Section~3.4).


The three independent isothermal elastic constants (\(C_{11}^{T}\), \(C_{12}^{T}\), \(C_{44}^{T}\)) were extracted from stress--strain relations via generalized Hooke's law. At each \emph{P--T} point, 20,480-atom supercells (16 × 16 × 16 unit cells) were equilibrated in the \emph{NPT} ensemble for 550 ps, averaging lattice parameters over the final 80\% of the trajectory. Finite strains (0 to ±0.30\% in 0.05\% increments) were applied, and stress responses were sampled in 550 ps \emph{NVT} production runs. Adiabatic constants for seismic velocities were obtained from their isothermal counterparts: \(C_{11}^{S}\) = \(C_{11}^{T}\) + \emph{αγTK\textsubscript{T}}, \(C_{12}^{S}\) = \(C_{12}^{T}\) + \emph{αγTK\textsubscript{T}}, \(C_{44}^{S}\) = \(C_{44}^{T}\), where \emph{α} is the thermal expansivity, \emph{γ} the Grüneisen parameter, and \emph{K\textsubscript{T}} the isothermal bulk modulus (\emph{αγT} ≡ \emph{C\textsubscript{P}}/\emph{C\textsubscript{V}} − 1). Heat capacities \emph{C\textsubscript{P}} and \emph{C\textsubscript{V}} were computed from numerical \emph{T} derivatives of the enthalpy (∂\emph{H}/∂\emph{T})\emph{\textsubscript{P}} and internal energy (∂\emph{U}/∂\emph{T})\emph{\textsubscript{V}} in \emph{NPT} and \emph{NVT} simulations, respectively. Voigt--Reuss--Hill averaging (VRH; \citealp{Hill1963}) yielded the adiabatic bulk (\emph{K\textsubscript{S}}) and shear (\emph{G\textsubscript{S}}) moduli, from which seismic velocities were calculated as


\begin{equation}
V_{\mathrm{P}} = \sqrt{(K_{S} + 4/3G_{S})/\rho},\qquad V_{\mathrm{S}} = \sqrt{G_{S}/\rho},
\tag{1}\label{eq:1}
\end{equation}


with \emph{ρ} the density at the given \emph{P--T} conditions. Figure 3a shows \emph{V}\textsubscript{P} and \emph{V}\textsubscript{S} at two representative mantle conditions, 12 GPa and 1,753 K, and 50 GPa and 2,142 K, along the 1,600 K-adiabat of \citet{Brown1981}. Our calculated velocities broadly agree with recent MLIP-MD simulations (\citealp{ZhangC2025}) and slightly exceed earlier small-cell (160 atom) DFT-MD estimates (\citealp{Kawai2015}), yet remain substantially higher than experimental measurements. At 12 GPa and 1,753 K, our calculated \emph{V}\textsubscript{S} (5.813 ± 0.001 km s\textsuperscript{-1}; mean ± 2 standard errors, \emph{n} = 10) exceeds the 300 K experimental value (5.25 ± 0.06 km s\textsuperscript{-1}) reported by \citet{Thomson2019} by more than 10\%. Because \emph{V}\textsubscript{S} decreases with \emph{T}, applying thermal corrections to the experimental data would further widen this discrepancy, reflecting a longstanding inconsistency between theory and experiment.


At 20,480 atoms, both finite-size effects and statistical uncertainties are negligible and cannot account for the offset. One proposed explanation is precursor elastic softening near the tetragonal--cubic phase transition (\citealp{ZhangC2025}). To test this hypothesis, we examined the \emph{T} dependence of \emph{K\textsubscript{S}}, \emph{G\textsubscript{S}}, and the derived seismic velocities at 50 GPa. Whereas \emph{K\textsubscript{S}} decreases smoothly from 432 GPa at 300 K to 399 GPa at 2,142 K, \emph{G\textsubscript{S}} exhibits strongly non-monotonic behavior, rising sharply from 30 GPa to a maximum of 241 GPa near 600--700 K before declining linearly to 215 GPa at higher \emph{T}. This behavior drives a large variation in \emph{V}\textsubscript{S} (2.5--7.0 km s\textsuperscript{-1}) but a comparatively modest change in \emph{V}\textsubscript{P} (9.8--12.4 km s\textsuperscript{-1}). The anomalously low \emph{G\textsubscript{S}} near 300 K reflects precursor softening adjacent to the phase transition. However, this regime lies well below typical mantle geotherms and therefore cannot explain the \textgreater10\% overestimation of \emph{V}\textsubscript{S} under ambient mantle conditions. Precursor softening is thus insufficient to reconcile theory and experiment.


Because DFT reproduces both \emph{V}\textsubscript{P} and the \emph{P}--\emph{ρ} relation of Dvm, the selective failure for \emph{V}\textsubscript{S} cannot reflect a breakdown of the theory. Finite-size effects, statistical uncertainty, and phase transitions are ruled out above. We are left with microstructure as the leading explanation. Experimental p-Dvm samples (\citealp{Greaux2019}; \citealp{Thomson2019}; \citealp{Zhou2025}) inevitably contain structural disorder absent from the ideal single crystals used in DFT calculations (\citealp{Kawai2015}; \citealp{Stixrude2007}). In disordered materials, shear deformation promotes nonaffine atomic rearrangements that strongly reduce \emph{G}, whereas \emph{K} is affected much less because atomic packing constraints inhibit such relaxations during compression (\citealp{Zaccone2014}). Quantifying this sensitivity is the focus of what follows.


\textbf{3.4. Polycrystalline velocity reduction}


Assuming elastic isotropy, we derive the bulk and shear moduli of each polycrystal from stress--strain relations using generalized Hooke's law. Thermoelastic properties were subsequently calculated following the methodology used for single-crystal Dvm (Section 3.3). Figure 3b shows \emph{ρ}, \emph{K\textsubscript{S}}, \emph{G\textsubscript{S}}, \emph{V}\textsubscript{P}, \emph{V}\textsubscript{S} as functions of \emph{Φ} (0--5.8 vol\%) at 50 GPa and 2,142 K. Density declines marginally by 2.7\% (4.74 to 4.61 g cm\textsuperscript{-3}), whereas elastic moduli soften markedly, particularly in shear: \emph{G\textsubscript{S}} drops by 37\% (213 to 136 GPa) compared with a 12\% reduction in \emph{K\textsubscript{S}} (401 to 353 GPa). This selective shear weakening propagates to seismic velocities: \emph{V}\textsubscript{S} decreases by 19.3\% (6.72 to 5.42 km s\textsuperscript{-1}), while \emph{V}\textsubscript{P} declines by only 10.6\% (12.03 to 10.76 km s\textsuperscript{-1}), buffered by the relative insensitivity of \emph{K\textsubscript{S}} to GB disorder.


The progressive reduction in \emph{G\textsubscript{S}} and \emph{V}\textsubscript{S} shifts polycrystalline properties from the ideal single-crystal limit toward experimental measurements. At \emph{Φ} = 5.8\%, \emph{V}\textsubscript{S} (5.42 km s\textsuperscript{-1}) falls below experimental extrapolations to these conditions (\citealp{Greaux2019}; \citealp{Thomson2019}; Fig. 3b), indicating that only a few volume percent of GB disorder can reconcile the discrepancy between single-crystal DFT predictions and polycrystalline Dvm measurements. These results establish nanoscale GB disorder as a key mechanism for reducing shear rigidity and seismic velocities in lower-mantle Dvm. To quantify this modulation, we modeled the aggregate as a two-phase composite comprising crystalline Dvm (\emph{ρ}\textsubscript{c}, \emph{K}\textsubscript{c}, \emph{G}\textsubscript{c}) and a disordered GB phase (\emph{ρ}\textsubscript{gb}, \emph{K}\textsubscript{gb}, \emph{G}\textsubscript{gb}) with volume fraction \emph{Φ}. Bulk density follows a linear volume-weighted mixture,


\begin{equation}
\rho_{\mathrm{bulk}} = (1 - \Phi)\rho_{\mathrm{c}} + \Phi\rho_{\mathrm{gb}},
\tag{2}\label{eq:2}
\end{equation}


which fits the data well \emph{R}\textsuperscript{2} = 0.996 with \emph{ρ}\textsubscript{gb} = 2.494 g cm\textsuperscript{-3}. Conversely, \emph{K\textsubscript{S}} and \emph{G\textsubscript{S}} decrease nonlinearly with \emph{Φ} (Fig. 3b). Comparison with classical homogenization bounds reveals that even modest GB disorder drives the aggregate toward the mechanically softest admissible limit. Using crystalline moduli \emph{K}\textsubscript{c} = 400.8 GPa and \emph{G}\textsubscript{c} = 213.9 GPa (Section 3.3), the Reuss (isostress) averages


\begin{equation}
\frac{1}{K_{\mathrm{R}}} = \frac{1 - \Phi}{K_{\mathrm{c}}} + \frac{\Phi}{K_{\mathrm{gb}}},\qquad \frac{1}{G_{\mathrm{R}}} = \frac{1 - \Phi}{G_{\mathrm{c}}} + \frac{\Phi}{G_{\mathrm{gb}}},
\tag{3}\label{eq:3}
\end{equation}


closely track the computed moduli (Fig. 3b, dashed lines). Linear regression of 1/\emph{K}\textsubscript{R} and 1/\emph{G}\textsubscript{R} against \emph{Φ} yields \emph{K}\textsubscript{gb} = 119.1 GPa and \emph{G}\textsubscript{gb} = 18.9 GPa (\emph{R}\textsuperscript{2} = 0.994 and 0.989, respectively). By contrast, the Voigt (isostrain) upper bounds,


\begin{equation}
K_{\mathrm{V}} = (1 - \Phi)K_{\mathrm{c}} + \Phi K_{\mathrm{gb}},\qquad G_{\mathrm{V}} = (1 - \Phi)G_{\mathrm{c}} + \Phi G_{\mathrm{gb}},
\tag{4}\label{eq:4}
\end{equation}


substantially overestimate the moduli. At \emph{Φ} = 5.8\%, \emph{G}\textsubscript{V} = 201.5 GPa exceeds the MLIP-MD value by 48\%, suggesting that aggregate elasticity is governed by GB compliance rather than strain averaging. To define the tightest physically admissible range, we evaluated the Hashin--Shtrikman (HS) bounds. Treating the compliant grain boundaries as inclusions within a crystalline matrix, the lower bounds are:


\begin{equation}
K_{\mathrm{HS}}^{-} = K_{\mathrm{gb}} + \frac{1 - \Phi}{\frac{1}{K_{\mathrm{c}} - K_{\mathrm{gb}}} + \frac{3\Phi}{3K_{\mathrm{gb}} + 4G_{\mathrm{gb}}}},\qquad G_{\mathrm{HS}}^{-} = G_{\mathrm{gb}} + \frac{1 - \Phi}{\frac{1}{G_{\mathrm{c}} - G_{\mathrm{gb}}} + \frac{6\Phi(K_{\mathrm{gb}} + 2G_{\mathrm{gb}})}{5G_{\mathrm{gb}}(3K_{\mathrm{gb}} + 4G_{\mathrm{gb}})}},
\tag{5}\label{eq:5}
\end{equation}


The HS upper bounds follow by interchanging matrix and inclusion phases (c ↔ gb). The HS lower bounds lie marginally above the Reuss limits but reproduce the MLIP-MD moduli to within 1.5\%. At \emph{Φ} = 5.8\%, \(K_{\mathrm{HS}}^{-}\) = 352.5 GPa versus 353.2 GPa (-0.2\%), and \(G_{\mathrm{HS}}^{-}\) = 133.6 GPa versus 135.7 GPa (-1.5\%). The corresponding HS upper bounds closely approximate the Voigt limits and substantially exceed the MLIP-MD results. This confirms that elasticity in disordered polycrystalline Dvm is governed by stress partitioning through a compliant, interconnected GB network. Using the HS lower-bound \emph{K} and \emph{G}, and Eq. (2) for \emph{ρ}, seismic velocities were calculated via Eq. (1). The resulting velocities match the MLIP-MD values to within 0.1 km s\textsuperscript{-1} (Fig. 3b). We note that although \emph{ρ}\textsubscript{gb}, \emph{K}\textsubscript{gb}, and \emph{G}\textsubscript{gb} are small relative to single-crystal Dvm and CaSiO\textsubscript{3} glasses (e.g., \citealp{Su2026}), they should be read as effective parameters describing the aggregate's sensitivity to intergranular disorder, not as intrinsic bulk-phase properties: extrapolated \textasciitilde17× beyond the simulated range (\emph{Φ} = 0--5.8 vol\%) to \emph{Φ} = 1, they are not directly comparable with bulk high-\emph{P} glass data.


\section*{4. Discussion}


In this section, we first assess the extent to which the disorder fraction \emph{Φ} can be constrained experimentally and discuss measurements that may help reconcile laboratory observations with theoretical predictions of p-Dvm elasticity. Here, \emph{Φ} denotes the total microstructural disorder expected in samples synthesized at high \emph{P--T}, encompassing both GB networks and possible residual amorphous material inherited from glass precursors---rather than the purely geometric GB fraction defined in Section 3.2. We then explore the implications of GB softening for lower-mantle seismic heterogeneity and grain size.


\textbf{4.1. Experimental constraints and uncertainties in \emph{Φ}}


Previous \emph{in situ} high \emph{P--T} velocity studies of p-Dvm synthesized from glass precursors (\citealp{Greaux2019}; \citealp{Thomson2019}; \citealp{Zhou2025}) did not directly quantify \emph{Φ}. Post-experiment characterization provides limited additional constraints because Dvm amorphizes upon decompression (\citealp{Irifune1994}), obscuring the high \emph{P--T} microstructure that might otherwise preserve evidence of disorder.


\citet{Greaux2019} reported the first \emph{in situ} sound velocities of p-Dvm under lower-mantle conditions, obtaining a zero-pressure shear modulus of \emph{G}\textsubscript{0} = 126 GPa, \textasciitilde26\% lower than theoretical predictions (\textasciitilde171 GPa; \citealp{Kawai2015}; \citealp{Stixrude2007}). They attributed this discrepancy to computational artifacts, including finite-size effects; however, our size-convergence tests rule out this explanation and instead indicate that GB disorder contributes to the reduced elasticity (Fig. 3b). Their inference of near-complete crystallization (\emph{Φ} ≈ 0) rested on two observations: (i) sharp XRD peaks with low background after 1 h annealing at 21 GPa and 1,300 K, and (ii) the absence of measurable \emph{V}\textsubscript{P} or \emph{V}\textsubscript{S} changes during successive heating cycles.


Although powder XRD is fundamental to phase identification, it is comparatively insensitive to structurally disordered domains. In silicate systems, amorphous content typically becomes detectable only above \textasciitilde10 vol\% (\citealp{Smith2018}), and this threshold is likely higher in LVP and DAC settings, where constrained sample geometries, high backgrounds, and limited angular coverage imposed by anvils suppress weak diffuse scattering from disordered material. A closely related challenge is the detection of incipient partial melting in laser-heated DAC experiments, where small melt fractions can modify elastic properties while remaining difficult to identify by XRD. Resolving such trace amounts often requires specialized multichannel collimators and dedicated diffuse scattering measurements (\citealp{Kim2020}).


Our simulated XRD patterns show this limitation directly: increasing \emph{Φ} from 0 to 5.8\% leaves Bragg peak positions and relative intensities essentially unchanged, producing only subtle peak broadening and weak diffuse scattering that would likely be obscured under experimental conditions. Importantly, p-Dvm with \emph{Φ} within this XRD-insensitive regime already reproduces the experimentally observed reductions in \emph{V}\textsubscript{P} and \emph{V}\textsubscript{S} relative to single-crystal DFT predictions (Fig. 3b). Published experiments illustrate the same point. \citet{Greaux2019} reported sharp Bragg peaks, suggesting \emph{Φ} ≈ 0, whereas \citet{Zhou2025} observed broader peaks indicative of higher \emph{Φ}. Despite these contrasting characteristics, both studies reported comparable velocities (\citealp{Zhou2025}, their Fig. 1). If peak sharpness quantitatively tracked the disorder fraction controlling elasticity, the sharper diffraction pattern of Gréaux et al. would be expected to yield elastic properties closer to single-crystal Dvm than those reported by Zhou et al. This apparent paradox suggests that peak sharpness is qualitatively, but not yet quantitatively, diagnostic of \emph{Φ}, pending systematic calibration (e.g., internal standards, diffuse scattering measurements).


The inference that invariant \emph{V}\textsubscript{P} and \emph{V}\textsubscript{S} during heating implies \emph{Φ} ≈ 0 may conflate kinetic stability with complete crystallization. It overlooks the possibility that a residual amorphous fraction reaches a kinetically arrested steady state during the initial high \emph{P--T} annealing cycle, such that subsequent heating produces negligible velocity changes. Similar behavior has been observed during perovskite ceramic sintering, where amorphous phases become trapped at GB triple junctions and are incorporated into the polycrystalline framework, limiting further microstructural evolution (\citealp{Choi2004}). In high-\emph{P--T} CaSiO\textsubscript{3} aggregates, GB disorder may likewise persist after annealing, potentially stabilized by stress gradients inherent to DAC and LVP experiments, and remain undetectable from inter-cycle variations in \emph{V}\textsubscript{P} or \emph{V}\textsubscript{S}.


Quantifying small \emph{Φ} values in p-Dvm requires direct sensitivity to local structural disorder rather than long-range periodicity or indirect proxies such as acoustic velocities. We propose three complementary approaches. First, total x-ray scattering combined with two-phase pair distribution function refinement can simultaneously model crystalline and disordered components in the measured structure factor \emph{S}(\textbf{q}), incorporating diffuse scattering overlooked in conventional Bragg analysis and potentially detecting amorphous fractions below standard diffraction limits (\citealp{Egami2003}; \citealp{Klug1974}). Second, Raman spectroscopy and imaging can identify disorder through broad Si--O bending (400--600 cm\textsuperscript{-1}) and stretching (800--1200 cm\textsuperscript{-1}) bands, in contrast to the sharp phonon modes of crystalline CaSiO\textsubscript{3} (\citealp{McMillan1984}). Third, thermodynamically driven segregation of incompatible elements to GBs (\citealp{Hiraga2004}) may provide a quench-stable compositional fingerprint accessible by \emph{ex situ} NanoSIMS (\citealp{Hao2016}), even after structural degradation during recovery. Without such constraints, the discrepancy between experimentally derived \emph{G} (and \emph{V}\textsubscript{S}) and DFT predictions retains an unquantified uncertainty arising from unresolved \emph{Φ}.


\section*{4.2. GB softening and the seismic signature of LLSVPs}


The origin of Earth's LLSVPs remains a major unresolved problem in deep-Earth geophysics. Whether they represent purely thermal anomalies or thermochemical reservoirs of recycled or primordial material has important implications for mantle dynamics, composition, and evolution. To evaluate the geophysical consequences of elastically softened p-Dvm, we calculated \emph{V}\textsubscript{P} and \emph{V}\textsubscript{S} for representative lower mantle assemblages using the Dvm elasticity derived here and, for comparison, single-crystal DFT elasticities from \citet{Kawai2015}. We considered two assemblages along a 1,600 K mantle adiabat (\citealp{Katsura2025}): (i) MORB, composed of 32 vol\% bridgmanite, 25 vol\% Dvm, 25 vol\% CF-phase, and 18 vol\% stishovite; and (ii) pyrolite, composed of 82 vol\% bridgmanite, 13 vol\% ferropericlase, and 5 vol\% Dvm. Calculated velocities were compared with the preliminary reference Earth model (PREM; \citealp{Dziewonski1981}) and with seismic observations of LLSVPs, which exhibit large negative \emph{V}\textsubscript{S} anomalies (2--3\%) and comparatively small \emph{V}\textsubscript{P} reductions (1\%) (\citealp{Garnero2016}).


Figs. 3c,d demonstrate that incorporating elastically softened p-Dvm alters predicted lower mantle seismic structure. Relative to models using single-crystal DFT elasticities, both pyrolite and MORB assemblages can exhibit systematically lower seismic velocities, with the effect concentrated overwhelmingly in shear. Whereas \emph{V}\textsubscript{P} decreases only modestly, \emph{V}\textsubscript{S} is strongly suppressed, reflecting the pronounced shear softening induced by disordered Dvm. Because MORB contains substantially more Dvm than pyrolite (25\% versus 5\%), the effect is amplified in basaltic assemblages, producing an LLSVP-like seismic signature. Relative to pyrolite, MORB calculated using the present Dvm elasticity exhibits \emph{V}\textsubscript{P} reductions of 1.9\% and \emph{V}\textsubscript{S} deficits of 6.2\% at 55 GPa. Relative to PREM, the same assemblage reproduces substantial shear-wave slowing (Δ\emph{V}\textsubscript{S} = -6.8\%) while maintaining comparatively small compressional anomalies (Δ\emph{V}\textsubscript{P} = -3.7\%). These calculations show that microstructural disorder in Dvm---if present at the grain sizes we simulate---could generate strong, selectively shear-weighted seismic anomalies in basalt-rich assemblages. We emphasize that this is a demonstration of mechanism and magnitude, not a claim that such fine-grained Dvm actually occurs within LLSVPs: as discussed below (Section 4.3), independent seismic attenuation evidence (\citealp{TalaveraSoza2025}) instead suggests LLSVPs are comparatively coarse-grained, implying small \emph{Φ} near the low-disorder end of the range explored here.


\section*{4.3. A lower bound on lower-mantle Dvm grain size}


Grain size \emph{d} is a fundamental microstructural parameter controlling the physical properties of polycrystalline mantle minerals (\citealp{Karato2008}), with the contribution of disordered regions to aggregate properties increasing as \emph{d} decreases. To first order, the fraction of structurally disordered GB scales as \emph{Φ} ≈ \emph{δ}/\emph{d}, where \emph{δ} is the width of the disordered interfacial region. From the structural maps in our simulations (Fig. 2a), we estimate \emph{δ} ≈ 1 nm. As shown in Fig. 3b, \emph{V}\textsubscript{S} is sensitive to \emph{Φ}, whereas \emph{V}\textsubscript{P} is comparatively buffered. Reproducing the large \emph{V}\textsubscript{S} reductions characteristic of LLSVPs while preserving their relatively modest \emph{V}\textsubscript{P} anomalies therefore requires \emph{Φ} to remain below \emph{Φ}\textsubscript{max} ≈ 1\%. Applying this scaling yields a minimum grain size,


\begin{equation}
d \gtrsim \frac{\delta}{\Phi_{\max}} \approx \frac{1\,\mathrm{nm}}{0.01} \approx 100\,\mathrm{nm}.
\tag{6}\label{eq:6}
\end{equation}


This mineral physics constraint represents the minimum \emph{d} required to avoid excessive velocity reductions associated with GB softening. Importantly, this bound is consistent with geodynamic considerations: models of grain size evolution in the convecting mantle indicate that background lower-mantle grain sizes typically evolve toward the micrometer scale (≫100 nm) through the competition between grain growth and dynamic recrystallization (e.g., \citealp{Dannberg2017}). At such coarse grain sizes, the fraction of mechanically active disordered interfaces becomes very small (\emph{Φ} ≪1\%), minimizing their contribution to aggregate elasticity and preserving ambient lower-mantle seismic velocities. Furthermore, this minimum \emph{d} constraint is compatible with recent seismic attenuation-based interpretations of LLSVP structure, which suggest that coarse-grained LLSVPs may coexist with finer-grained regions in the surrounding circum-Pacific mantle (\citealp{TalaveraSoza2025}).


Two limitations define priorities for future work. First, our simulations assume pure CaSiO\textsubscript{3}, whereas natural Dvm contains Na, K, Fe, Al, and Ti impurities (\citealp{Tschauner2021}) that may modify its elasticity. Second, our grains (Fig. 2a), though near the practical limit of atomistic simulation, remain orders of magnitude below expected mantle grain sizes (\citealp{Dannberg2017}). Meeting both challenges will require MLIPs spanning multicomponent chemical spaces and methods reaching larger microstructural scales. Nevertheless, our results show that MLIP-based simulations already capture complex GB networks at near-DFT accuracy, linking mineral-scale microstructure to seismic observations of the deep Earth.


\textbf{Acknowledgments}


This work was supported by the National Natural Science Foundation of China (nos. 42572036 and 42225202), the Natural Science Foundation of Hubei Province (no. 2025AFB454), the National Key R\&D Program of China (no. 2023YFF0804100), and the Deep Earth Probe and Mineral Resources Exploration--National Science and Technology Major Project of China (no. 2025ZD1005508).


\textbf{Open research}


All data supporting the findings of this study are available from the corresponding author on reasonable request. Electronic structure calculations were performed with VASP (v5.4.4; \citealp{Kresse1996}). Molecular dynamics simulations were performed using LAMMPS (v2Aug2023; \citealp{Thompson2022}) and GPUMD (v5.0; \citealp{Fan2022}). Machine learning interatomic potentials were developed with DeePMD-kit (v2.2.10; \citealp{Wang2018}), DP-GEN (v0.10.2; \citealp{Zhang2020}), and GPUMD-NEP (v5.0; \citealp{Fan2021}; \citealp{Fan2022}). Polycrystal models were generated with Atomsk (v0.13.1; \citealp{Hirel2015}) and visualized with OVITO (v3.14.1; \citealp{Stukowski2010}).


\textbf{Conflict of interest}


The authors declare no conflicts of interest relevant to this study.


\clearpage
\begin{center}\includegraphics[width=0.85\linewidth,height=0.64\textheight,keepaspectratio]{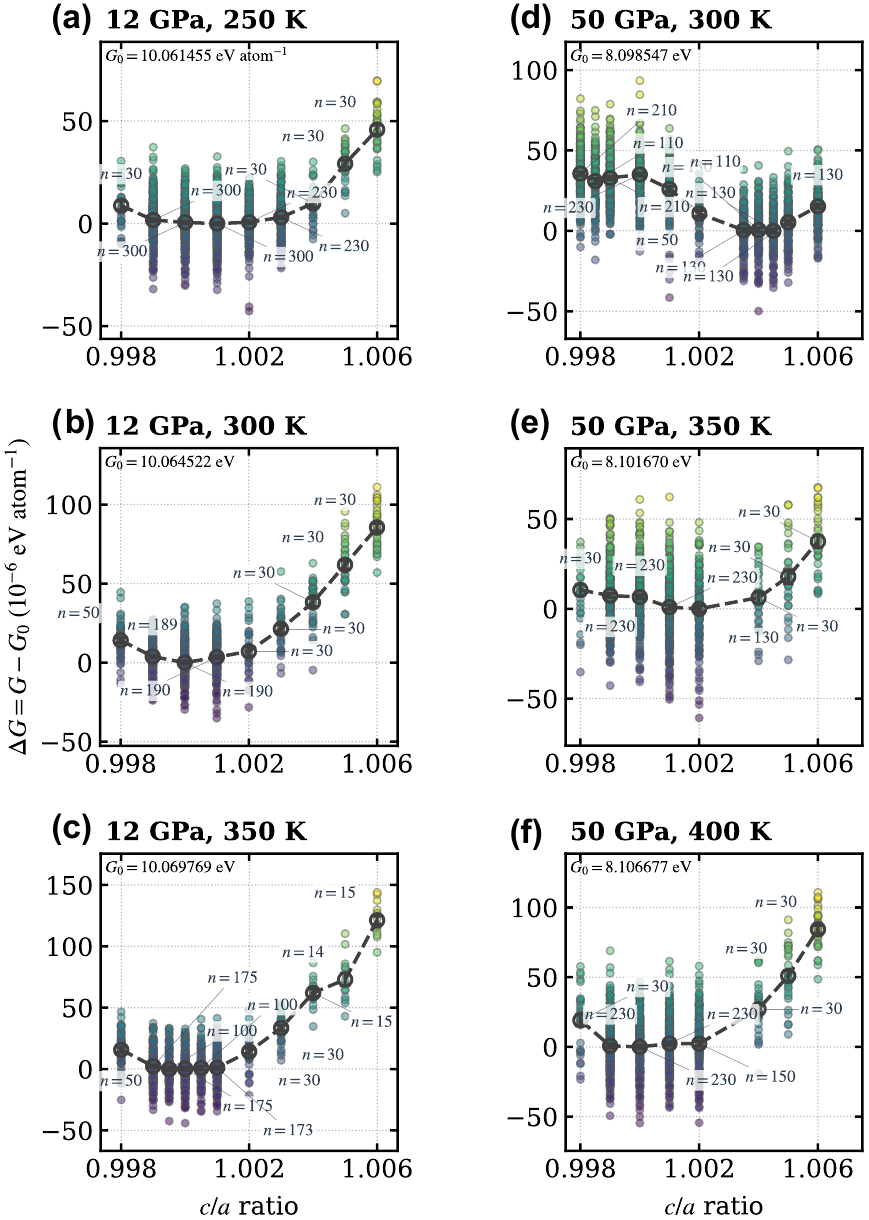}\end{center}
\textbf{Fig. 1: Free-energy landscape of CaSiO\textsubscript{3} perovskite across the tetragonal--cubic regime.} Gibbs free energy (\emph{G}) computed via thermodynamic integration is plotted against the axial ratio (\emph{c}/\emph{a}) at 12 and 50 GPa, spanning 250--400 K. Small colored markers show individual NE-TI runs; large circles connected by dashed lines are the ensemble-averaged free energies (\(\overline{G}\)), with annotations indicating the number of statistically independent samples (\emph{n} ≤ 300) per \emph{c}/\emph{a} value. Energies are plotted relative to panel-specific reference values \emph{G}\textsubscript{0} (annotated at top left of each panel) to highlight fine-scale variations.


\clearpage
\begin{center}\includegraphics[width=\linewidth,height=0.62\textheight,keepaspectratio]{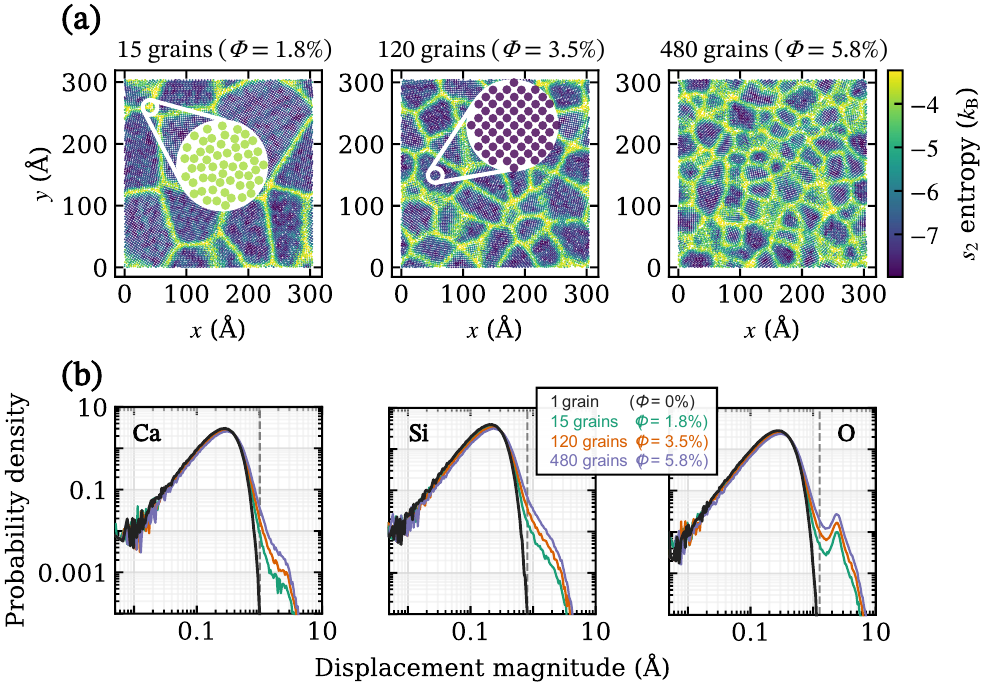}\end{center}
\textbf{Fig. 2: Structural and dynamical fingerprints of grain-boundary disorder in polycrystalline davemaoite.} \textbf{(a)} Maps of two-body entropy (\emph{s}\textsubscript{2}; \citealp{Piaggi2017}) at 50 GPa and 2,142 K for 15-, 120-, and 480-grain models; insets are schematic illustrations of characteristic local atomic environments. Progressive grain refinement drives boundary interconnection, with ordered interiors (\emph{s}\textsubscript{2} ≈ -7.5 \emph{k}\textsubscript{B}, purple) distinguished from disordered grain boundaries and amorphous pockets (\emph{s}\textsubscript{2} ≈ -4.5 \emph{k}\textsubscript{B}, yellow). \textbf{(b)} Probability densities of atomic displacement magnitudes for Ca, Si, and O over 1 ns at 50 GPa and 2,142 K, shown for the single crystal and the three polycrystals. The single-crystal baseline (black) is dominated by thermal vibrations below the species-specific threshold displacement \emph{d}\textsubscript{0} (vertical dashed lines), whereas the polycrystals develop pronounced non-Gaussian tails extending beyond \emph{d}\textsubscript{0}, reflecting enhanced atomic mobility at grain boundaries. The fraction of atoms with displacements exceeding \emph{d}\textsubscript{0} defines the disordered volume fraction \emph{Φ}, which increases systematically with grain refinement from 1.8 vol\% (15 grains) to 3.5 vol\% (120 grains) and 5.8 vol\% (480 grains).


\clearpage
\begin{center}\includegraphics[width=\linewidth,height=0.60\textheight,keepaspectratio]{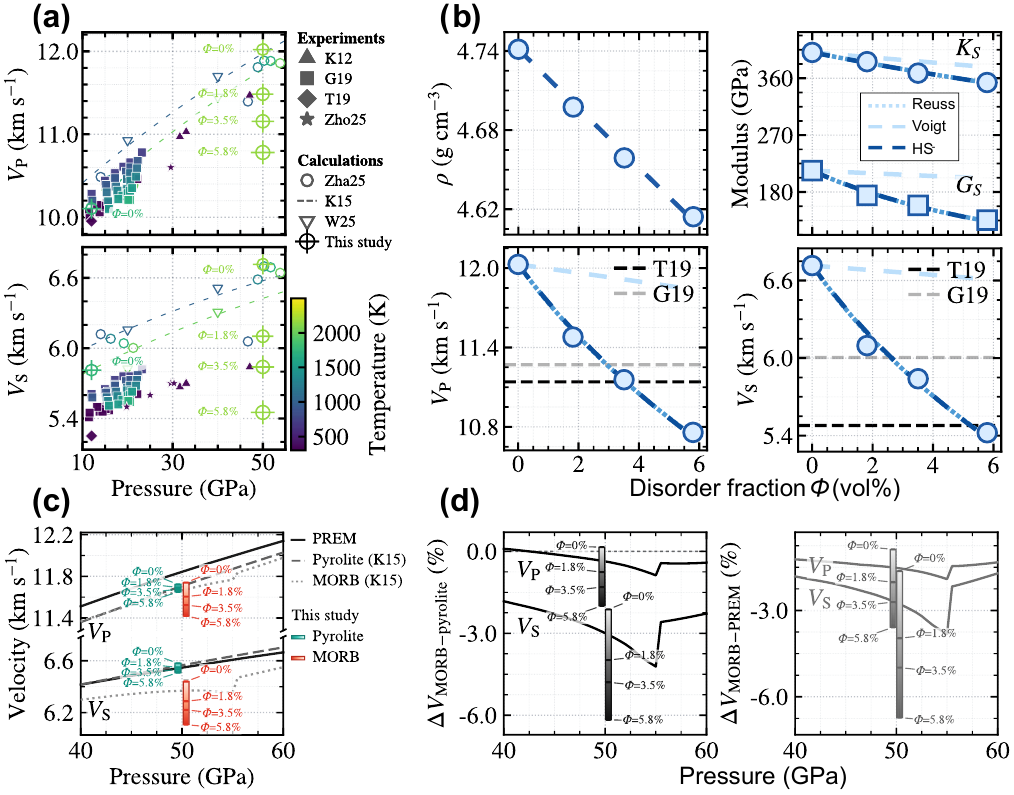}\end{center}
\textbf{Fig. 3: Elastic softening of polycrystalline davemaoite.} \textbf{(a)} Single-crystal DFT-predicted compressional (\emph{V}\textsubscript{P}) and shear (\emph{V}\textsubscript{S}) velocities (open symbols: Zha25, \citealp{ZhangC2025}; W25, \citealp{Wan2025}; dash-dotted line: K15, \citealp{Kawai2015}) systematically exceed polycrystalline measurements (filled symbols: K12, \citealp{Kudo2012}; G19, \citealp{Greaux2019}; T19, \citealp{Thomson2019}; Zho25, \citealp{Zhou2025}) across mantle \emph{P--T} conditions (color denotes temperature). Circled crosses show this study's results at \emph{Φ} = 0--5.8 vol\%. \textbf{(b)} At 50 GPa and 2,142 K, increasing disorder (\emph{Φ} = 0--5.8 vol\%) monotonically decreases density (\emph{ρ}), bulk (\emph{K\textsubscript{S}}) and shear (\emph{G\textsubscript{S}}) moduli, and seismic velocities. Shear softening dominates: \emph{V}\textsubscript{S} drops \textasciitilde20\% while \emph{V}\textsubscript{P} declines only \textasciitilde10\%, matching the experimental bounds (T19, G19; horizontal dashed lines). Colored lines show Reuss (dotted), Voigt (light dashed), and Hashin--Shtrikman (HS; dark dashed) fits to the MLIP-MD results (symbols). \textbf{(c)} Aggregate velocities for pyrolite and MORB using our davemaoite elasticity, and K15, along a 1,600-K mantle adiabat, compared with PREM (\citealp{Dziewonski1981}). \textbf{(d)} Anomaly profiles (Δ\emph{V}\textsubscript{MORB--pyrolite} and Δ\emph{V}\textsubscript{MORB--PREM}) show that grain-boundary disorder drives strong \emph{V}\textsubscript{S} reductions with weaker \emph{V}\textsubscript{P} anomalies. This \emph{V}\textsubscript{P}--\emph{V}\textsubscript{S} decoupling captures the LLSVP seismic fingerprint. The \textasciitilde55 GPa kink marks the stishovite--CaCl\textsubscript{2} silica transition proposed by \citet{Yang2014}. All other mantle parameters follow \citet{Zhou2025} and references therein.
\clearpage

\end{document}